# A SYSTEMATIC LITERATURE REVIEW OF CLOUD COMPUTING IN eHEALTH

Yan Hu and Guohua Bai

Department of Creative Technologies, Blekinge Institute of Technology, Sweden

## ABSTRACT

*Cloud computing in eHealthis an emerging area for only few years. There needs to identify the state of the art and pinpoint challenges and possible directions for researchers and applications developers. Based on this need, we have conducted a systematic review of cloud computing in eHealth. We searched ACM Digital Library, IEEE Xplore, Inspec, ISI Web of Science and Springer as well as relevant open-access journals for relevant articles. A total of 237 studies were first searched, of which 44 papers met the Include Criteria. The studies identified three types of studied areas about cloud computing in eHealth, namely (1) cloud-based eHealth framework design (n=13);(2) applications of cloud computing (n=17); and (3) security or privacy control mechanisms of healthcare data in the cloud (n=14). Most of the studies in the review were about designs and concept-proof. Only very few studies have evaluated their research in the real world, which may indicate that the application of cloud computing in eHealth is still very immature. However, our presented review could pinpoint that a hybrid cloud platform with mixed access control and security protection mechanisms will be a main research area for developing citizen centred home-based healthcare applications.*

## 1. INTRODUCTION

This review examines existing researches on cloud-based eHealth solutions. The main goal is to identify the state of the art in this area and pinpoint challenges and possible directions for researchers and applications developers based on the current literatures. Though this studymay not able to specify thebenefits of using cloud technology in eHealth due to the progress in the area so far is made mostly in designs and concept-proof, not in real use context, we do, however, have identified some better ways of using cloud computing in eHealth.

eHealthis defined as "the cost-effective and secure use of information and communications technologies in support of health and health-related fields, including healthcare services, health surveillance, health education, knowledge and research".[1]The goal of eHealth is to improve the cooperation and coordination of healthcare, in order to improve the quality of care and reduce the cost of care at the same time.

Cloud computing is a new technology which has emerged in the last five years. According to the definition by NIST, cloud computing is "a model can provide distributed, rapidly provisioned and configurable computing resources (such as servers, storage, applications, networks and other services), which are on-demand, rapid elastic and measured, to whom have network connections".[2]Because of the obvious scalability, flexibility and availability at low cost of cloud services, there is a rapid trend of adopting cloud computing among enterprises or health-related areas in the last few years.

## 2. METHODS

A systematic literature review requires a comprehensive and unbiased coverage of searched literatures. To maximise the coverage of our searched literatures, we started by identifying some of the most used alternative words/concepts and synonyms in the research questions.[3] We conducted first a manual search in the areas of related areas such as computer science and healthcare. The selected Databases are ACM Digital Library, IEEE Xplore, Inspec, ISI Web of Science and Springer. In order to cover more broad scope, open-access journals in the relevant areas were also included. We did not limit the publication year, since cloud computing was proposed only in the last five years. After general study of the related areas, the language of the papers was limited to English. The following search string was used to search the above mentioned databases:

(Cloud)AND (eHealth OR "electronic health" OR e-health)

The search string could be modified slightly when searching in different databases, since they have different rules for search strings. Our first search by the search string in all the mentioned databases produced 237 articles. In order to focus on the most relevant literatures, we conducted a primary evaluation based on reading the abstracts of all selected articles. The evaluation is based on the criteria described in Table 1. The inclusion criteria are applied independently for each author to select the relevant articles. This evaluation selected 44 articles for our thorough study, and all are included in the reference list.

Table 1. Include and exclude criteria

| Include criteria | Exclude criteria |
|---|---|
| • Directly or indirectly related to both eHealth and cloud technology.<br>• Cloud-based eHealth frameworks design.<br>• Cloud computing solutions applied in healthcare.<br>• Security and privacy mechanisms of healthcare data in cloud.<br>• Written in English | • Irrelevant to study of the cloud or eHealth.<br>• Conceptual methods or cognitive introductions.<br>• Review papers.<br>• Business analysis reports.<br>• Not written in English. |

The quality of each paper was assessed by two authors mainly based on the Jovell and Navarro-Rubio system for classification from score 9 to 1 [4]. Guidelines for performing systematic literature reviews in related subject area[5] were followed for technology related issues. After include/exclude criteria and quality assessment criteria, a chosen set of papers was available for the data extraction process. In order to avoid the bias of subjective preference, we applied the method by which one researcher extracted the data and another checked the extraction. The Citation and Bibliography tool – Zotero – was used to manage all the extracted articles.

## 3. RESULTS

After the steps of searching, evaluation and alternate reviews, 44 articles were finally selected from the total of 237 articles found from the first search. We believe the selected 44 articles can cover the basic view of the studied area of cloud computing in eHealth. Since both eHealth and cloud computing are emerging areas, the results of this study can offer the researchers up-to-date perspectives for their research.

We found 19 countries where research articles were published on eHealth cloud computing. The largest number of articles were produced in the USA (n=14), followed by EU countries (n=11). We found relatively few papers in this new area produced in developing countries such as China (n=3), India (n=2) and UAE (n=3) (see Table 2). All papers we searched were published after 2010, and this may indicate that research in cloud computing for healthcare is still an emerging area. Figure 1 shows the number of papers published across years in our searched results.

Table 2. Countries conducting cloud-based eHealth researches

| Country | No. of papers | References |
|---|---|---|
| USA | 14 | 7,11,13,17,18,20,25,29,30,32,35 |
| Canada | 4 | ,41,43,45 |
| Australia | 3 | 23,38,39,40 |
| UK | 3 | 20,44,48 |
| UAE | 3 | 10,14,26 |
| China | 3 | 8,21,33 |
| Germany | 2 | 34,36,47 |
| India | 2 | 46,49 |
| France | 1 | 9,24 |
| Croatia | 1 | 42 |
| Italy | 1 | 12 |
| Spain | 1 | 15 |
| Sweden | 1 | 27 |
| Netherland | 1 | 19 |
| Bangladesh | 1 | 28 |
| Puerto Rico | 1 | 37 |
| Egypt | 1 | 31 |
| Brazil | 1 | 16 |
| | | 6 |

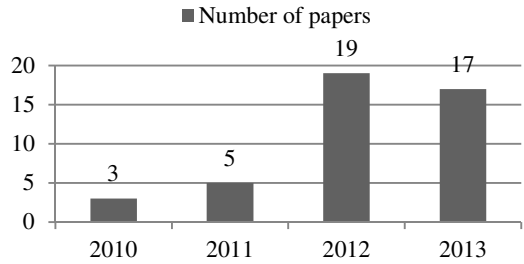

Figure 1. Number of studies across years

In the following, we present some important findings from our study. The topics discussed in the reviewed articles are quite broad in relation to the use of cloud computing in eHealth domain. Generally, the topics can be classified into three categories: 1) Cloud-based eHealth framework design (n=13); 2) Applications of cloud computing in eHealth (n=17); and 3) Security or privacy control mechanisms of healthcare data in cloud (n=14). The distribution of the above topics is shown in Figure 2.

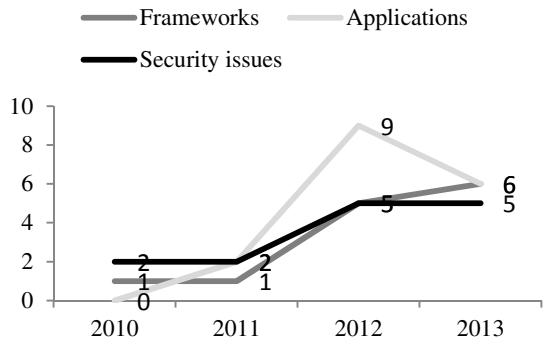

Figure 2.Research type distributions.

## 3.1 Discussion of the topics in the reviewed articles

### 3.1.1Cloud-based eHealth framework design

Since one of the most significant advantages of cloud computing is its huge data storage capacity, six papers proposed cloud-based frameworks for healthcare data sharing. One of the pioneers in this area, Rolim et al. [6] designed a framework for data collection by using sensors attached to medical equipment, and the collected data can be directly stored in a cloud, which can be accessed by authorised medical staff. Some studies[7][8][9]  proposed a national level framework for eHealth based on cloud models. For example, Patra et al. [9] argued especially that their cloud-based solution on a national level would provide a cost effective way in dealing with patient information for rural areas. By encouraging people in rural areas to upload their personal healthcare information to the health cloud, the care providers can provide them with more correct healthcare services, such as remote diagnosis, supervision and emergency calls.

Other studies related to this category of framework design are more specific as regards application areas, such as the design of a Virtual Research Environment by both Simth et al.[10] and Regola et al.[11]; patients' self-management by Martinovic et al.[12]; transition or standardisation of data stored in different EHR or PHR systems by Coats et al.[13]and Ekonomou et al.[14]; and designing a secure EHR framework [15][16][17][18].

### 3.1.2.Applications of cloud computing

High accessibility, availability and reliability make cloud computing a better solution for healthcare interoperability problems. Papers in this category mostly applied cloud technology for healthcare data sharing, processing and management, and can be categorised based on three types of cloud platforms, namely, public cloud, private cloud and hybrid cloud.

Six papers presented their eHealth applications by using or testing in public clouds such as Google App Engine [19], Windows Azure[20][21] and Amazon EC2[22][23][24].The application[22] of Wooten et al. provided a patients-to-patients support and information sharing within patients community. The solution proposed by Benharref et al.[21] used the mobiles of seniors to send the patients' data automatically to the cloud, and the patients themselves could decide with whom to share the data. Mohammed et al.[23]designed a Health Cloud Exchange (HCX) system which shares healthcare records between services and consumers with some privacy controls.

For applications based on the private cloud, Bahga et al.[25] presented an achievement of sematic interoperability between different kinds of healthcare data, while DACRA[26] built a platform for interoperability on the syntax level. Vilaplana et al.[27]used queuing theory as the basic means to model the performance of an eHealth system based on the private cloud. Van Gorp et al.[28]applied virtualisation techniques to allow patients themselves to build their own lifelong PHRs. The PHR can then be shared with other stakeholders who are authorised and interested. Wu et al.[29]proposed an approach to EHR data schema composition with a broker based access control. In order to reduce the cost of adopting EHRs, HP published a cloud-based platform called Fusion[30]for securely managing and sharing healthcare information on large scale. Other studies also used the private cloud to integrate the EHR systems with other systems like healthcare billing system[31] and national law system[32].

Gul et al.[33] and Chen et al.[34]proposed a shared EHRs system based on hybrid cloud. In the proposed application of Chen et al.[34], the patient's medical data are stored both in a hospital's private cloud and public healthcare cloud. A mechanism is set up to make sure that the owners of the medical records can decide when their records should be protected in normal or emergency situation. Dixon et al.[35]implemented a community cloud-based exchange of clinical data between two disparate health care providers, which was mainly used by chronic disease healthcare.

### 3.1.3.Security or privacy control mechanisms

Healthcare data require protection for high security and privacy. Access control, an effective method to protect data, is widely used in many studies. Liu et al. [36] applied an identity based encryption (IBE) system in access control of PHR, and this identity-based cryptography system can reduce the complexity of key management. Attribute based Encryption (ABE) is one of the most preferable encryption schemes used in cloud computing. For example, Fakhrul et al.[37] implemented Cipher text-Policy ABE in a security manager module to make it act as an administrative person; ESPAC[38] and Narayan et al.[39] proposed a patient-centric ABE access control scheme; and Aljumah et al.[40]designed an emergency mobile access to PHR cloud-based ABE.

Three researches[41][42][43] mixed ABE and IBE to identify access on different levels (normal and emergency), which can handle more complex situations than a single scheme. Role based access control is based on ABE, which is an automatic procedure for authenticating healthcare user information and allocating corresponding role to guarantee all associated operations. Tong et al.[44]introduced a Cloud-based Privacy-aware Role Based Access Control model for controllability, traceability of data and authorised access to healthcare resources. Sharma et al.[45]developed an advanced role-based scheme called task based control to determine whether access should be granted to a healthcare cloud.

Besides access control, several security protection techniques (Trusted Virtual domains[46], Watermarking method[47], Secure index implementation[48] and secret-sharing schemes[49]) were also introduced to maintain the high security and privacy of healthcare clouds.

## 4. DISCUSSIONS

The presented review shows that cloud computing technology could be applied in several areas of the eHealth domain. The majority of studies introduced cloud computing technology as possible solutions for achieving eHealth interoperability. Although worldwide it is acknowledged that ICT technologies, such as cloud computing, can improve healthcare quality, most papers in this review are from developed countries. The present review does show, however, increasing studies from other developing countries.

Almost all the studies suggest that due to the huge amount of patient health data, especially in the case of daily care, the cloud's big data storage service provides a better way to store these data. The data can be shared among hospitals and third party research institutions or other healthcare organisations even on a national level. The huge data storage capacity of the cloud would help the development of big data mining in healthcare, as well as diagnosis and treatment. Pay-as-you-go mode of the cloud has significant economic strength, reducing cost for all healthcare organisations which would like to use cloud-based services.

The patients-centric healthcare model will be a future trend where patients are active participants in their own healthcare. Some studies presented cloud-based patients-centric healthcare applications by the users-centric feature of cloud computing. This will not only encourage healthcare receivers to be involved in their own healthcare, but also the cloud-based healthcare platform will provide a technical solution and a social network. In addition, healthcare receivers' participation constitutes an efficient healthcare education in terms of patient's self-management.

High accessibility and availability of the cloud could help the healthcare data stored in the cloud to be accessed at anytime and anywhere in the world. If healthcare receivers could make parts of their healthcare data in public cloud open, which means that data can be "freely used, reused and redistributed by anyone – subject only, at most, to the requirement to attribute and sharelike."[50]. When the open data become available in the public cloud, it can be processed by remote services, such as medical systems in hospital, clinic decision support systems, expert systems, or distributed to other medical personnel. Around one third of the studies also show that the security and privacy gaps of healthcare data in the cloud could be solved by access control encryption schemes and security protection techniques. This would make it possible to move current server-client based eHealth services to cloud-based eHealth services and make more contribution to improve the current healthcare by high-technologies.

The present review also noted some challenges to using cloud computing in eHealth. Healthcare data contain sensitive information, and dealing with sensitive data in the cloud could lead to some legal issues. Besides, it is important to select cloud providers carefully to guarantee the confidentiality of healthcare data.

Based on the review, we could find that a hybrid cloud model which contains access controls and security protection techniques would be a reliable solution. The EHRs in the hospitals and other healthcare centres could keep their data in private clouds, while patients' daily self-management data could be published in a confident public cloud. Patients as the owner of their health data should decide who can access their data and the conditions for sharing.

## 5. LIMITATIONS

The present review has certain limitations. The most obvious one is the external validity or generalizability. Since cloud computing is a relatively new technology, the number of published works for our review topic is not so large. In order to have a wider spectrum of studies, the selection rate is slightly higher. Since all the databases we used for searching articles contain some intelligent search techniques, we did not create as many synonyms of "eHealth" as we may do. This may have caused some data loss. Since cloud computing came after 2010, and most of the studies in this review are conference papers with concept-proof-designs. Only very few studies havebeen evaluated in the real world or tested by some technical experts.

## 6. CONCLUSION

Research on applying cloud computing technology to eHealth is in its early stages; most researchers have presented ideas without real-world cases validation. The obvious features of cloud computing technology provide more reasons to adopt cloud computing in sharing and managing health information. The main purpose of our review is to identify some challenges and feasible cloud-based solutions which can be applied in eHealth. The current review suggests that with the unique superiority of the cloud in big data storage and processing ability, a hybrid cloud platform with mixed access control and security protection mechanisms will be a main research area for developing citizen centred home-based healthcare system.